\documentclass[lettersize,journal]{IEEEtran}
\usepackage{titlesec} 
\usepackage[english]{babel}
\usepackage[font=small]{caption} 
\usepackage{footnote}
\usepackage{multicol}
\usepackage{tabularx}
\RequirePackage{booktabs}  
\usepackage{graphicx}
\usepackage[colorlinks=true, allcolors=blue]{hyperref}
\usepackage{algorithm}
\usepackage{amsthm,amsmath,amssymb,lipsum}
\usepackage{mathrsfs}
\usepackage{multirow}
\usepackage{enumitem}
\usepackage{amsthm}
\usepackage{diagbox} 
\usepackage{float} 
\usepackage{subcaption}
\usepackage{algpseudocode}
\usepackage{xcolor}
\usepackage[backend=biber,style=ieee,url=false,doi=false]{biblatex}

\begin{document}

\title{Maritime Communication in Evaporation Duct Environment with Ship Trajectory Optimization}
\author{
Ruifeng Gao, {\em Member, IEEE},
Hao Zhang,
Jue Wang, {\em Member, IEEE},
Ye Li, {\em Member, IEEE},\\
Yingdong Hu, {\em Member, IEEE},
Qiuming Zhu, {\em Senior Member, IEEE},
Shu Sun, {\em Senior Member, IEEE},\\
Meixia Tao, {\em Fellow, IEEE}

\thanks{Ruifeng Gao is with the School of Transportation and Civil Engineering, Nantong University, Nantong 226019, China (e-mail: grf@ntu.edu.cn).}

\thanks{Hao Zhang, Jue Wang, Ye Li and Yingdong Hu are with the School of Information Science and Technology, Nantong University, Nantong 226019, China (e-mail: 2330310050@stmail.ntu.edu.cn; \{wangjue,yeli,huyd\}@ntu.edu.cn).}


\thanks{Qiuming Zhu is with the College of Electronic and Information Engineering, Nanjing University of Aeronautics and Astronautics, Nanjing 211106, China (e-mail: zhuqiuming@nuaa.edu.cn).}

\thanks{Shu Sun and Meixia Tao are with the School of Information Science and Electronic Engineering, Shanghai Jiao Tong University, Shanghai 200240, China (e-mail: \{shusun,mxtao\}@sjtu.edu.cn).}}

\maketitle

\begin{abstract}

In maritime wireless networks, the evaporation duct effect has been known as a preferable condition for long-range transmissions. However, how to effectively utilize the duct effect for efficient communication design is still open for investigation. In this paper, we consider a typical scenario of ship-to-shore data transmission, where a ship collects data from multiple oceanographic buoys, sails from one to another, and transmits the collected data back to a terrestrial base station during its voyage. A novel framework, which exploits priori information of the channel gain map in the presence of evaporation duct, is proposed to minimize the data transmission time and the sailing time by optimizing the ship's trajectory. To this end, a multi-objective optimization problem is formulated and is further solved by a dynamic population PSO-integrated NSGA-II algorithm. Through simulations, it is demonstrated that, compared to the benchmark scheme which ignores useful information of the evaporation duct, the proposed scheme can effectively reduce both the data transmission time and the sailing time. 


\end{abstract}


\IEEEpeerreviewmaketitle

\section{Introduction}
With the rapid growth of marine economic activities, there is an increasing demand for reliable and high-speed maritime communication to support diverse applications such as life monitoring, offshore oil exploration, aquaculture, and weather observation \cite{9023458}. These applications often rely on the maritime Internet of Things (IoT), where sensor data must be collected and transmitted over long distances. Current maritime communication systems primarily rely on satellite communications \cite{10363686}, airborne platform communications \cite{8960465}, and shore-based communications \cite{9165199}. Satellite communications provide global coverage and support broadband services, but face several challenges, including high latency and significant service costs \cite{9344715}. Airborne platform communications provide deployment flexibility, but their reliability and surveillance capabilities are limited due to susceptibility to mobility-induced instability \cite{9771332} and constrained onboard energy reserves \cite{8721541}, particularly in large-scale deployments. In contrast, shore-based base stations (BSs) leverage mature terrestrial wireless technologies to provide reliable, high-capacity, low-cost services to near-shore maritime users \cite{9147146}. In such systems, ships can act as mobile data relay, gathering information from dispersed sensor nodes and forwarding it to terrestrial centers via wireless links \cite{9939173}. However, their effective coverage is fundamentally limited by the radio transmission range over the sea, posing a major challenge for long-distance communications \cite{8528349,sun2025landtoship58ghz}.
 
Evaporation ducts present a valuable propagation approach to extend this range \cite{8624585}. This phenomenon is a result of seawater evaporation and the sea-sky temperature difference, leading to a sharp decrease in relative humidity with altitude near the sea surface. The resulting humidity gradient alters the atmospheric refractive index, creating a ducting layer in which electromagnetic waves experience super-refraction. This trapping effect enables guided propagation between the sea surface and the duct boundary, where most of the electromagnetic wave energy is confined. As a result, path loss is greatly reduced \cite{5741881}, which is significantly different from traditional propagation in free space. In particular, electromagnetic waves within the duct can propagate beyond the line of sight (LoS) limit, often extending over 40 kilometers, thus allowing beyond line of sight (BLoS) communication \cite{6917399}. Notably, a 78 km microwave link has been established between the Australian mainland and the Great Barrier Reef \cite{5191066}, and field experiments in the South China Sea have demonstrated BLoS communication over 100 km \cite{9785409}. Given this extended coverage, using ships as mobile data collectors and leveraging evaporation duct propagation offers a promising solution for maritime data transmission.

Evaporation duct formation and persistence are primarily governed by the air–sea temperature difference (ASTD), wind speed, and relative humidity  \cite{yang2022regional}. Under unstable stratification, when $\text{ASTD}$ is less than $0^{\circ}\mathrm{C}$, stronger winds and drier air generally intensify the duct \cite{shi2015new}, whereas rainfall and frontal passages tend to erode it. The duct also exhibits pronounced diurnal and seasonal variability and can persist for extended periods. Observational studies further show marked regional contrasts in occurrence: in mid- to high-latitude seas the incidence is around 50\%, while in low-latitude it often exceeds 80\% \cite{zhao2020research}. Consequently, the evaporation duct can be regarded as a widespread marine feature. Motivated by this, by leveraging duct-assisted propagation, we investigate a ship-to-shore communication scenario in which a ship gathers data from ocean buoys and transmits it to a shore-based BS during its voyage.

Effective communication via evaporation ducts requires accurate modeling of electromagnetic wave propagation. Since analytical modeling of the evaporation duct channel is not available, existing research can be categorized into two types of approaches: numerical modeling and artificial neural network methods. On the one hand, traditional numerical models involve ray optics (RO) \cite{Mehrnia2018RayTM} or parabolic equation (PE)-based calculations \cite{sirkova2006parabolic}. Due to the high computational complexity of the RO method and the adoption of the split-step Fourier method \cite{sirkova2012brief}, the PE method has become the primary approach and is widely used. On the other hand, neural networks, trained on realistic measurement data, can be employed to model and analyze practical evaporation duct wave propagation \cite{10328712}. To enable effective communication, the channel characteristics under evaporation duct conditions have been investigated. The authors in \cite{8668540} employ PE to predict the characteristics of wireless propagation within offshore evaporation ducts. \cite{9831108} integrates three-dimensional subsection ray tracing with the shooting-and-bouncing ray to model both propagation and scattering effects. The authors in \cite{10036458} further propose a path loss prediction approach based on gated recurrent unit neural networks. \cite{9813428} uses least squares estimation to predict model coefficients, demonstrating better performance in predicting path loss within evaporation ducts.

With the predicted path loss, maritime communications can be optimized. In \cite{8721541}, the correlation between pre-known ship lane and static channel information is exploited to optimize resource allocation for multiuser maritime communication. More recently, a pre-established database of path loss, commonly referred to as a channel gain map (CGM), has emerged as a promising technique for capturing environment-specific propagation characteristics and enabling environment-aware communication \cite{10891198}. Based on the CGM, users can obtain path loss using their positional data, facilitating system design. In \cite{10371362}, such a channel map is constructed to support communication design in cognitive satellite–UAV networks. Specifically, path loss is retrieved from the map via user and UAV position queries, significantly reducing system overhead required for communication design.

In this paper, we propose a novel framework for optimizing ship-to-shore data transmission in evaporation duct environment. In particular, we jointly minimize transmission and sailing time by optimizing the ship's trajectory, where the spatial distribution information of path loss, in the presence of evaporation duct, is pre-stored in a CGM. To the best of our knowledge, this work is the first study of maritime communications with optimization of the trajectory of ships considering the evaporation duct effect. The main contributions of this work are summarized as follows.

\begin{itemize}
  \item We mathematically formulate the problem of efficient maritime data collection and ship-to-shore communication, which incorporates the effects of the evaporation duct, and also accounts for practical ship navigation constraints. To facilitate trajectory design, we develop a simplified yet practical ship motion model tailored for maritime communication scenarios.
  \item We exploit CGM of the evaporation duct channel to facilitate the ship trajectory and communication design. Discrepancy between the resolution of CGM and the ship's sailing trajectory is addressed, for which a resolution alignment method is proposed to ensure calculation accuracy.
  \item An optimization framework is proposed to determine the ship's trajectory, leveraging the duct CGM to reduce the time required for both communication and sailing. A \underline{\textbf{D}}ynamic \underline{\textbf{P}}opulation 
  \underline{\textbf{P}}SO-\underline{\textbf{I}}ntegrated \underline{\textbf{NSGA-II}} (DPPI-NSGA-II) algorithm is developed to solve the optimization problem efficiently.

\end{itemize}

The remainder of this paper is organized as follows. Section~\ref{section:A} introduces the system model and formulates the problem. Section~\ref{section:F} discretizes the problem and proposes the trajectory-CGM resolution alignment scheme. Section~\ref{section:C} presents the proposed optimization algorithm for trajectory and transmission design. Section~\ref{section:D} presents the simulation results along with discussion. Finally, Section~\ref{section:E} concludes the paper.

\textit{Notations:} $|\cdot|$ denotes the absolute value of a scalar. $\lceil x \rceil$ denotes the ceiling operation, which returns the smallest integer greater than or equal to $x$, and $\lfloor x \rfloor$ denotes the floor operation, which returns the largest integer less than or equal to $x$. $||\textbf{x}||$ denotes the Euclidean norm of vector $\textbf{x}$. $\emptyset$ represents the empty set. $\mathbb{R}^{x \times y}$ denotes the space of real-valued matrices of size $x \times y$. $O(\cdot)$ denotes the Big-O notation that describes the asymptotic computational complexity. $\text{U}[0, 1]$ denotes a uniform distribution between 0 and 1.

\section{System Model and Problem Formulation}
\label{section:A}

\begin{figure}[t!]
\centering
\includegraphics[width=0.43\textwidth]{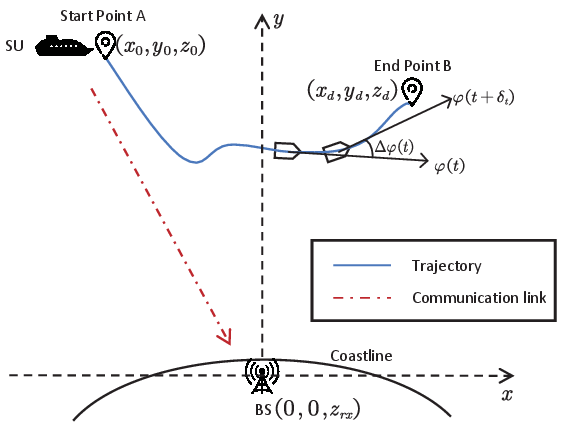}
\captionsetup{font={small,stretch=1.25},justification=raggedright}
\caption{System Model.}
\label{fig:myfigure} 
\end{figure}

We consider a maritime communication system consisting of a ship user (SU) and a shore-based BS. As shown in Fig.~\ref{fig:myfigure}, the SU sails from a start location $A$ to an end location $B$. During the voyage, the SU is required to transmit its collected data, with a total amount of $D$ bits, to the BS as soon as possible. For simplicity and for the reason that we focus on ship trajectory optimization, it is assumed that both the BS and the SU are equipped with a single antenna for communication. In practice, if directional antenna or multi-antenna beamforming is adopted, our analysis is still valid by simply setting the antenna gain pattern to conform to actual situation. 

The objective is to minimize both the data transmission time and the SU's sailing time by optimizing the SU’s trajectory, exploiting information of the evaporation duct environment at sea. In the following, we first describe the coordinate system, the ship motion model, as well as the signal model. Then, an optimization problem is formulated for the considered data transmission task. 

\subsection{Coordinate System and Ship Motion Model}
A three-dimensional Cartesian coordinate system is established to describe the scenario. We assume that the height of the transmit antenna is $z_{tx}$ and the height of the receive antenna is $z_{rx}$. The coordinates of the start point $A$ and end point $B$ are defined as $\mathbf{c}_A = (x_0, y_0, z_0)$ and $\mathbf{c}_B = (x_d, y_d, z_d)$, respectively. The BS is located at $(0, 0, z_{rx})$. The location of the SU at time $t$ is expressed as
\begin{equation}
    \mathbf{c}_s(t) = \big(x_s(t), y_s(t), z_s(t)\big), \quad 0 \leq t \leq T, \label{equ1}
\end{equation}
where $T$ denotes the maximum allowable sailing time. The velocity of the SU at time $t$ is denoted as $v(t)$, and its velocity vector angle, measured with respect to the $x$-axis, is denoted as $\varphi(t)$. Consequently, the velocity components of the SU in the $x$- and $y$-directions are given by
\begin{equation}
\begin{aligned}
    v_x(t) &= v(t) \cos[\varphi(t)], \quad 0 \leq t \leq T, \\
    v_y(t) &= v(t) \sin[\varphi(t)], \quad 0 \leq t \leq T,
\end{aligned}
\end{equation}
where $\varphi(t) \in[\varphi_l,\varphi_u]$, with $\varphi_l$ and $\varphi_u$ denoting the lower and upper bounds of $\varphi(t)$ respectively.

The ship's maneuverability is limited due to physical constraints. For the velocity directions between any two adjacent time instants, the steering angle must satisfy the following constraint
\begin{equation}
   \Delta \varphi (t) \triangleq |\varphi(t+\delta_t)-\varphi(t)|\leq \Delta_{\varphi_{max}}(\delta_t), \quad 0 \leq t \leq T, \label{3}
\end{equation}
where $\delta_t$ denotes the interval between two consecutive time instants, $\Delta_{\varphi_{max}}(\delta_t)$ denotes the maximum allowable change in the velocity vector angle of $\delta_t$ interval. 

\subsection{Signal model}

At time instant $t$, the received signal at the BS is
\begin{equation}
    y(t) = h(t)x(t) + n(t), \label{euq5}
\end{equation}
where $h(t)$ is the channel between the SU and the BS, $x(t)$ is the transmit signal, and $n(t)$ denotes additive white Gaussian noise. As a widely-adopted simplification in sparse scattering environments, we consider only large-scale fading when describing $h(t)$, which is thereby location-dependent \cite{8721541}. Besides, we assume constant transmit power, such that
\begin{equation}
    P_t = |x(t)|^2,
\end{equation}
and the receive signal power $P_r(t)$ is expressed as
\begin{equation}
    P_r (t)= |h(t)|^2 P_t = L_p(\mathbf{c}_s(t)) G_t G_r P_t, \label{equ8}
\end{equation}
where $L_p(\mathbf{c}_s(t))$ denotes the path loss between the SU and the BS, $G_t$ denotes the transmit antenna gain, and $G_r$ denotes the receive antenna gain.

\begin{figure}[h]
\centering
\includegraphics[width=0.47\textwidth]{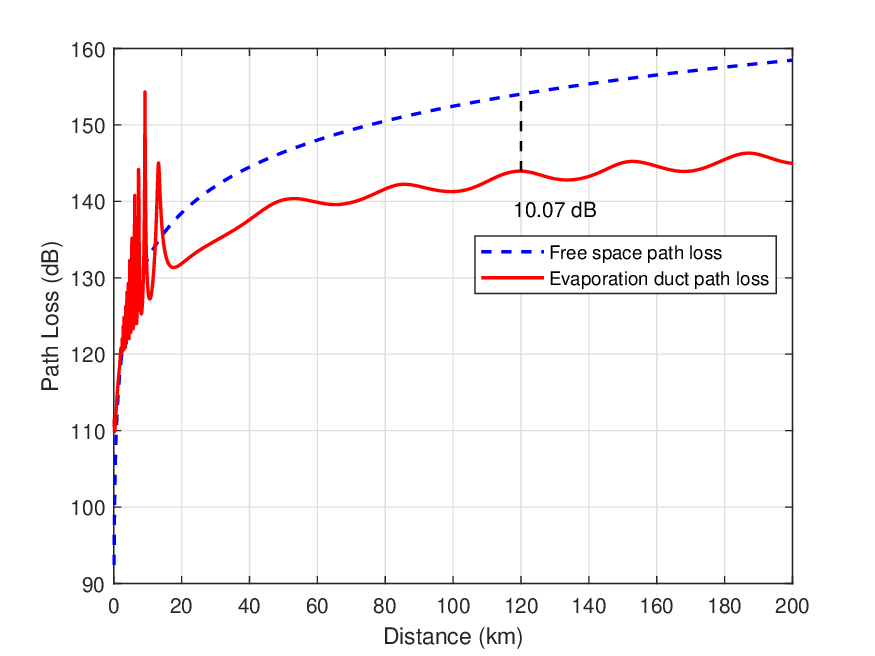}
\captionsetup{font={small,stretch=1.25},justification=raggedright}
\caption{Comparison of path loss in free space and evaporation duct environments. The evaporation duct path loss is reproduced based on Figure 3 in \cite{ouyang2023study}.}
\label{fig:fig3}
\end{figure}

The path loss of the evaporation duct in the marine environment is different from that in free space. Fig.~\ref{fig:fig3} shows an example for the path loss at carrier frequency of 10 GHz, where the receive antenna is located at a height of 18.3 m, the transmit antenna is located at a height of 25 m, and the height of the evaporation duct is 40 m. Unlike the free-space scenario, where path loss consistently increases with distance, the path loss in an evaporation duct environment exhibits oscillatory variations as the distance increases. Moreover, path loss within the evaporation duct environment is typically lower than that in free space, particularly at longer distances. For instance, at a distance of 120 kilometers, the difference in path loss between these two environments can reach approximately 10.07 dB. Therefore, more efficient communication design is expected if the information of duct channel is properly exploited.


\subsection{Problem formulation}
Based on the ship motion model and the signal model, we now formulate the optimization problem. Let $T_1$ denote the data transmission time and $T_2$ denote the sailing time. Let $\mathbf{w} = \{\mathbf{c}_s(t), t \in [0, T_2]\}$ denote the trajectory of the SU. The optimization problem is given by
\begin{flalign}
    \textbf{(P1)}:  \mathop{\text{min}}\limits_{\mathbf{w}} & \quad T_1, T_2  \\
    \text{s.t.} & \quad T_1 \leq T_2, \tag{7a} \label{7a} \\
    & \quad  T_2 \leq T, \tag{7b} \label{7b} \\
    & \quad \int_{0}^{T_1} R(t) dt \geq D, 0\leq t \leq T_1,\tag{7c} \label{7c} \\
    & \quad \mathbf{c}_s(0) = \mathbf{c}_A = (x_0, y_0,z_0), \tag{7d} \label{7d} \\
    & \quad \mathbf{c}_s(T_2) = \mathbf{c}_B = (x_d, y_d,z_d), \tag{7e} \label{7e} \\
    & \quad |\varphi(t+\delta_t)-\varphi(t)|\leq \Delta_{\varphi_{max}}(\delta_t),0 \leq t \leq T, \tag{7f} \label{7f}
\end{flalign}
where constraint (\ref{7a}) ensures that the collected data must be fully transmitted to the BS, before the SU reaches the end location B. Constraint (\ref{7b}) represents the maximum sailing time. Constraint (\ref{7c}) indicates that the total amount of data to be transmitted is $D$. Constraints (\ref{7d}) and (\ref{7e}) specify the ship's initial and final location conditions, respectively. Constraint (\ref{7f}) provides practical restriction on the maximum steering angle between two consecutive time instants.

According to (\ref{euq5})-(\ref{equ8}), the maximum transmit rate $R(t)$ at any time $t$ in (\ref{7c}) can be expressed as
\begin{equation}
    R(t) = C=B\log_2\left(1+\frac{L_p(\mathbf{c}_s(t))G_{t}G_{r}P_{t}}{n_0B}\right), \label{13}
\end{equation}
where $B$ is the channel bandwidth, $n_0$ is the noise power spectral density. We assume that the path loss $L_p(\mathbf{c}_s(t))$ can be retrieved from a pre-established CGM based on the SU’s location $\mathbf{c}_s(t)$. In the following, we will present the problem transformation using CGM.

It is worth mentioning that the practicality of CGM depends on both the temporal resolution of meteorological inputs and the intrinsic variability of the evaporation duct. CGMs can be constructed across time scales by combining reanalysis and forecasts for climatological baselines, in-situ platforms for timely profiles, and refractivity-from-clutter for near-real-time, spatially resolved fields \cite{hu2023observed, qiu2023analysis, mai2020new}. Moreover, the duct exhibits pronounced diurnal and seasonal cycles driven by the air–sea temperature difference, near-surface wind, and humidity, observations indicate that 4$-$6 h refreshes are adequate under stable conditions, whereas 1 h updates are advisable during disturbed periods \cite{compaleo2021refractivity, yang2022regional}.

\section{Problem Transformation Using CGM}
\label{section:F}
In this section, we first discretize and simplify \textbf{(P1)} using the CGM. To ensure consistency between the CGM resolution and the SU’s sailing trajectory, we then propose an alignment scheme to match their granularities.

\subsection{Problem Discretization}
We assume that the evaporation duct CGM is discretized into three-dimensional grids, each with a horizontal width of $\Delta d$ and a vertical height of $\Delta h$. Within each grid, the channel gain is assumed constant. The coordinates of each grid $s(x,y,z)$ can be expressed as
\begin{equation}
    s(x,y,z)=(x\Delta d,y\Delta d,z\Delta h),x\in I_N,y\in J_N,z\in K_N,
\end{equation}
where $I_N=\left\{-N_{le},-N_{le}+1,\cdots,N_{le}-1,N_{le}\right\}$, $J_N=\left\{0,1,\cdots,N_{le}-1,N_{le}\right\}$, $K_N=\left\{0,1,\cdots, N_{ve}-1, N_{ve}\right\}$. The channel gain at any grid $s(x,y,z)$ can be expressed as $\overline{L_p}(s(x,y,z))$. At time $t$, if the SU is located in the grid $s(x,y,z)$, we have $\overline{L_p}(\Upsilon(\mathbf{c}_s(t)))=\overline{L_p}(s(x,y,z))$, where $\Upsilon(\mathbf{c}_s(t))$ is a function that identifies the grid containing $\mathbf{c}_s(t)$.


It is worth mentioning that the sizes of $\Delta h$ and $\Delta d$ must be carefully selected according to the actual environment. Unlike the relatively stable channel characteristics in long-distance horizontal propagation, variations in the heights of transmit and receive antennas significantly impact the large-scale channel gain \cite{9813428}. Specifically, even a difference of a few meters in antenna height can lead to path loss variations of up to 10 dB. Given that the evaporation duct height ranges from 0 to 40 m, the vertical grid resolution $\Delta h$ should be set very small to reduce the path loss difference between vertical grids. Furthermore, considering that path loss exhibits slower horizontal variation compared to vertical changes within the evaporation duct \cite{10130128}, the horizontal grid resolution can be coarser relative to the vertical resolution. Accordingly, $\Delta d$ should be chosen to balance accuracy against computational and storage costs.

With the CGM of the evaporation duct, we replace $L_p(\mathbf{c}_s(t))$ in (\ref{13}) with the data-driven expression $\overline{L_p}(\Upsilon(\mathbf{c}_s(t)))$. However, since $\overline{L_p}(\Upsilon(\mathbf{c}_s(t)))$ lacks a closed-form analytical expression, problem \textbf{(P1)} becomes difficult to solve directly. While heuristic algorithms are commonly used for such problems, they face challenges when the solution space is continuous. To address this, we discretize the entire sailing duration into multiple timeslots, each with a duration of $\Delta t$. Let $M_1$ denote the number of timeslots allocated for data transmission, $M_2$ denote the number of timeslots required for the SU to reach the end location B, and $M = \lceil \frac{T}{\Delta t}\rceil$ denote the total number of timeslots corresponding to the maximum allowable sailing time. Accordingly, the SU's trajectory can be expressed as $\mathbf{w} = \{\mathbf{c}_s(0), \mathbf{c}_s(1), \ldots, \mathbf{c}_s(M_2)\}$. We then rewrite (\ref{equ1}) and (\ref{3}) as follows

\begin{equation}
    \mathbf{c}_s(i) = (x_s(i),y_s(i),z_s(i)),0 \leq i \leq  M,
\end{equation}
\begin{equation}
   |\varphi(i+1)-\varphi(i)|\leq\Delta_{\varphi_{max}}(\Delta t),0 \leq i \leq M-1, \label{15}
\end{equation}
where $\mathbf{c}_s(i)$ represents the location of the SU at the $i$-th timeslot. Accordingly, (\ref{7c}) in \textbf{(P1)} is modified as follows

\begin{equation}
  \sum_{i=1}^{M_1} R(i)\Delta t  \geq D. \label{16}
\end{equation}

By discretizing the sailing time into equal-length timeslots, the continuous-time objectives $T_1$ and $T_2$ are transformed to corresponding timeslot indices $M_1$ and $M_2$. Problem \textbf{(P1)} can be reformulated as
\begin{flalign}
    \textbf{(P2)}:  \mathop{\text{min}}\limits_{\mathbf{w}} & \quad M_1, M_2  \\
    \text{s.t.} & \quad M_1 \leq M_2, \tag{13a} \label{13a} \\
    & \quad  M_2 \leq M, \tag{13b}\label{13b} \\
    & \quad \sum_{i=1}^{M_1} R(i)\Delta t  \geq D, \tag{13c}\label{13c} \\
    & \quad \mathbf{c}_s(0) = (x_0,y_0,z_{0}), \tag{13d}\label{13d} \\
    & \quad \mathbf{c}_s(M_2) = (x_d,y_d,z_{d}), \tag{13e}\label{13e} \\
    & \quad |\varphi(i+1)-\varphi(i)|\leq \Delta_{\varphi_{max}}(\Delta t),0 \leq i \leq M-1, \tag{13f} \label{13f}
\end{flalign}
where $R(i)$ can be expressed as
\begin{equation}
    R(i)=B\log_2\left(1+\frac{\overline{L_p}(\Upsilon(\mathbf{c}_s(i)))G_{t}G_{r}P_{t}}{n_0B}\right),
\end{equation}
where $\overline{L_p}(\Upsilon(\mathbf{c}_s(i)))$ depends on the location of the SU at the beginning of the $i$-th timeslot. However, the SU may cross multiple grids in the CGM within one timeslot, leading to the calculation errors of $R(i)\Delta t$. Additionally, both data transmission completion and arrival at the destination may occur within the final timeslot, further affecting calculation accuracy. To address these issues, it is essential to align the resolution of the SU's trajectory with that of the CGM.

\subsection{Trajectory-CGM Resolution Alignment}
We divide each trajectory segment within a single timeslot into multiple sub-timeslots. Assuming that the size of a sub-timeslot is $\delta t$, then one timeslot can be divided into $m$ sub-timeslots, where $m= \frac{\Delta t}{\delta t}$. The location of the SU at the $n$-th sub-timeslot of the $i$-th timeslot can be expressed as
\begin{equation}
    \mathbf{c}_s(i,n) = \left(\frac{\mathbf{c}_s(i+1)-\mathbf{c}_s(i)}{m}\right)n + \mathbf{c}_s^0(i),0 \leq n \leq m.
\end{equation}
The channel gain at this location is $\overline{L_p}(\Upsilon(\mathbf{c}_s(i,n)))$, and the transmit rate can be expressed as
\begin{equation}
     R\left(i,n \right)= B\log_2\left(1+\frac{\overline{L_p}(\Upsilon(\mathbf{c}_s(i,n)))G_{t}G_{r}P_{t}}{n_0B}\right).
\end{equation}





Assume the data transmission is completed within the $n_1$-th sub-timeslot in the $M_1$-th timeslot. The transmission time before the $n_1$-th sub-timeslot is $(M_1-1)\Delta t+(n_1-1)\delta t$, and the data that has been transmitted is $\sum_{i=1}^{M_1-1} R(i)\Delta t + \sum_{n=0}^{n_1-1} R(M_1,n)\delta t$. Then the data that needs to be transmitted in the $n_1$-th sub-timeslot is $D-\sum_{i=1}^{M_1-1} R(i)\Delta t - \sum_{n=0}^{n_1-1} R(M_1,n)\delta t$, so the transmission time in the $n_1$-th sub-timeslot is
\begin{equation}
\frac{D-\sum_{i=1}^{M_1-1} R(i)\Delta t - \sum_{n=0}^{n_1-1} R\left(M_1,n\right)\delta t}{R\left(M_1,{n_1}\right)}.
\end{equation}

Therefore, the modified $\widetilde{M_1}$ can be expressed as
\begin{equation}
\begin{aligned}
 \widetilde{M_1}= &M_1-1+(n_1-1)\frac{\delta t}{\Delta t}\\ 
 &+\frac{D-\sum_{i=1}^{M_1-1} R(i)\Delta t - \sum_{n=0}^{n_1-1} R\left(M_1,n\right)\delta t}{R\left(M_1,{n_1}\right)\Delta t}. \label{22}
 \end{aligned}
\end{equation}

Since the SU arrives within the $M_2$-th timeslot, the accumulated sailing time before this timeslot is $(M_2 - 1)\Delta t$. At the beginning of the $M_2$-th timeslot, the distance between the ship and the destination is $||\mathbf{c}_B-\mathbf{c}_s(M_2)||$. We assume that the sailing speed is $v$, the sailing time in timeslot $M_2$ is $\frac{||\mathbf{c}_B-\mathbf{c}_s(M_2)||}{v}$. The modified $\widetilde{M_2}$ can be expressed as
\begin{equation}
     \widetilde{M_2}= M_2 - 1 + \frac{||\mathbf{c}_B-\mathbf{c}_s(M_2)||}{v\Delta t}. \label{23}
\end{equation}

According to (\ref{22}), constraint (\ref{13c}) is given by
\begin{equation}
\begin{aligned}
    \sum_{j=1}^{\lfloor \frac{\widetilde{M_1}\Delta t}{\delta t}\rfloor} R(j)\delta t 
    + &R\left(\lfloor \frac{\widetilde{M_1}\Delta t}{\delta t}\rfloor+1\right)\\
    &\left(\widetilde{M_1}\Delta t -\lfloor \frac{\widetilde{M_1}\Delta t}{\delta t}\rfloor \delta t\right) \geq D.
\end{aligned}
\end{equation}

Thus \textbf{(P2)} can be rewritten as follows
\begin{flalign}
\textbf{(P3)}:
    \mathop{\text{min}}\limits_{\mathbf{w}} & \quad \widetilde{M_1}, \widetilde{M_2}  \\
    \text{s.t.} & \quad \widetilde{M_1} \leq \widetilde{M_2}, \tag{21a} \label{25a} \\
    & \quad \widetilde{M_2} \leq M, \tag{21b}\label{25b} \\
    & \quad R\left(\lfloor \frac{\widetilde{M_1}\Delta t}{\delta t}\rfloor+1\right)\left(\widetilde{M_1}\Delta t -\lfloor \frac{\widetilde{M_1}\Delta t}{\delta t}\rfloor \delta t\right) \notag \\
    & \quad \hspace{0cm}+ \sum_{i=1}^{\lfloor \frac{\widetilde{M_1}\Delta t}{\delta t}\rfloor} R(i)\delta t \geq D, \tag{21c}\label{25c} \\
    & \quad \mathbf{c}_s(0) = (x_0, y_0,z_{0}), \tag{21d}\label{25d} \\
    & \quad \mathbf{c}_s(\widetilde{M_2}) = (x_d,y_d,z_{d}), \tag{21e}\label{25e} \\
    & \quad |\varphi(i+1)-\varphi(i)|\leq \Delta_{\varphi_{max}}(\Delta t), 0 \leq i \leq \lceil M \rceil-1. \tag{21f}\label{25f}
\end{flalign}

However, \textbf{(P3)} is a multi-objective nonlinear programming problem that is hard to solve directly. Next, we propose an evolutionary algorithm that combines NSGA-II and PSO to tackle this problem.

\section{Optimization Algorithm for Trajectory and Transmission Design}
\label{section:C}
Multi-objective evolutionary algorithms demonstrate strong robustness and adaptability in solving multi-objective nonlinear programming problems, achieving consistent performance across many scenarios. Next, we introduce the NSGA-II algorithm and the PSO algorithm respectively and propose the DPPI-NSGA-II algorithm to solve \textbf{(P3)}.

\subsection{NSGA-II Algorithm}
NSGA-II is a fast multi-objective genetic algorithm developed as an enhancement of the NSGA. The key innovations of NSGA-II include a fast non-dominated sorting approach to reduce computational complexity, an elitism strategy to preserve high-quality solutions, and a crowding distance mechanism to ensure diversity without manual parameter tuning \cite{6786471}. These improvements enable NSGA-II to achieve faster convergence and better Pareto front coverage. However, NSGA-II relies on crossover and mutation operations to explore the solution space. In some scenarios, especially when the solution space is large, its local search capability may be insufficient, resulting in falling into the local optimum \cite{ma2023comprehensive}.

\subsection{Particle Swarm Optimization Algorithm}
PSO is an optimization algorithm inspired by swarm intelligence. Each particle, representing a potential solution to the problem, is characterized by two properties: position and velocity. The particles interact by sharing information and exchanging insights about the best positions they have discovered \cite{9783074}. Each particle updates its position and velocity according to the individual optimal position $p_{best}$ and the global optimal position $g_{best}$.

At each generation $\alpha$, particles update their velocity and position using the following equations \cite{juneja2016particle}
\begin{align}  
    v^{\alpha+1} &= \omega v^{\alpha} + c_1 r_1 (p_{best} - p^{\alpha}) + c_2 r_2 (g_{best} - p^{\alpha}), \label{26}\\
    p^{\alpha+1} &= p^{\alpha} + v^{\alpha+1}, \label{27}
\end{align}
where $v^{\alpha}$ and $p^{\alpha}$ denote the velocity and position of the particle of generation $\alpha$, $\omega$ is the inertia weight, $c_1$ is the cognitive learning factor, $c_2$ is the social learning factor, and $r_1, r_2 \sim U(0,1)$ are two random numbers.

PSO can provide better local search capability by adjusting the size of the inertia weight $\omega$. Combining the advantages of PSO's local search and NSGA-II's global search, it is possible to achieve a more balanced search method in \textbf{(P3)}.
\subsection{DPPI-NSGA-II Algorithm}
To address the local search limitations of NSGA-II and improve convergence efficiency, we propose the DPPI-NSGA-II algorithm by integrating the NSGA-II and PSO algorithms. The proposed algorithm consists of two main components.
\begin{itemize}
    \item We enhance NSGA-II by introducing constraint-based initialization to improve population diversity and convergence. Additionally, we dynamically adjust the solution space dimension by truncating redundant parameters based on the optimized sailing time, thereby reducing the computational complexity of genetic operations.
    \item The output from the modified NSGA-II serves as the input to the PSO algorithm. The PSO algorithm leverages the non-dominated solutions from NSGA-II to guide the swarm’s search process, effectively mitigating the risk of premature convergence to local optima.
\end{itemize}
 
 The details of the proposed DPPI-NSGA-II algorithm are outlined as follows.
 
\textbf{Step 1: Individual and population initialization.} We set the randomly generated decision variables $\Phi$ from \textbf{(P3)} as the first generation. The population is then initialized by grouping $N_P$ individuals, with half being randomly generated based on the maximum angle constraint, while the remaining half are generated according to the steering angle constraint of adjacent timeslots. The $\alpha$-th generation of the population is denoted as $Q^\alpha$.

\textbf{Step 2: Fitness values calculation and non-dominated sorting.} The individual \textbf{$\boldsymbol{\gamma}_k$} is evaluated using two fitness functions, $f_k^1$ and $f_k^2$. They are obtained by adding the penalty function to the corresponding objective function (\ref{22}) and (\ref{23}). The penalty functions are set as the violations of constraints (\ref{25a}), (\ref{25b}), and (\ref{25f}). The specific formulation is as 
\begin{equation}
\begin{aligned}
 f_k^1 = &\widetilde{M_1}+\max\left(\widetilde{M_1}-\widetilde{M_2},0\right)\\ 
 &+\iota \times \sum_{i=0}^{\lceil \widetilde{M_2} \rceil} \max\left( |\varphi(i+1) - \varphi(i)| - \varphi_{\text{max}}, 0 \right), \label{28}
 \end{aligned}
\end{equation}
\begin{equation}
\begin{aligned}
 f_k^2 = &\widetilde{M_2}+\max\left(\widetilde{M_2}-\frac{T}{\Delta t},0\right)\\ 
 &+\iota \times \sum_{i=0}^{\lceil \widetilde{M_2} \rceil} \max\left( |\varphi(i+1) - \varphi(i)| - \varphi_{\text{max}}, 0 \right), \label{29}
 \end{aligned}
\end{equation}
where $k$ refers to the individual \textbf{$\boldsymbol{\gamma}_k$}. $\iota$ is the penalty coefficient and is used to balance the trade-off between constraint violations and the objective function. 
After computing the fitness values of each individual, we determine the non-dominance level by sorting based on these values. Specifically, if $ \forall i \in \{1,2\}, (f_a^i \leq f_b^i) \: \& \: \exists j \in \{1,2\}, (f_a^j < f_b^j)$, then individual \textbf{$\boldsymbol{\gamma}_a$} is considered to have a higher non-dominance level than \textbf{$\boldsymbol{\gamma}_b$}, otherwise, they are assigned the same non-dominance level. 

For individuals with the same dominance level, we further rank them based on their crowding distance. The crowding distance $\ell_k$ of individual \textbf{$\boldsymbol{\gamma}_k$} is computed as
\begin{equation}
 \ell_k = \sum_{j=1}^{2}\frac{f_{k+1}^j-f_{k-1}^j}{f^j_{max}-f^j_{min}}, \label{30}
\end{equation}
where $f_{k+1}^j$ and $f_{k-1}^j$ represent the fitness values of the $k+1$-th and $k-1$-th individuals, respectively. $f^j_{max}$ and $f^j_{min}$ denote the maximum and minimum fitness values in the population, respectively. All individuals are sorted in ascending order based on their non-domination level and, in the case of ties, by their crowding distance.

\textbf{Step 3: Crossover and mutation.} We select the top $pd$ individuals from the population $Q^\alpha$ of the $\alpha$-th generation as the candidate population $Q_t$ for crossover and mutation. During the crossover, two parents \textbf{$\boldsymbol{\gamma}_a$}, \textbf{$\boldsymbol{\gamma}_b$} are randomly selected from $Q_t$, and then two offspring are generated as follows 
\begin{equation}
\begin{aligned}
 & \boldsymbol{\gamma}_a^{\alpha+1} = 0.5[(1+\boldsymbol{\beta})\boldsymbol{\gamma}_a^\alpha+(1-\boldsymbol{\beta})\boldsymbol{\gamma}_b^\alpha], \\
 & \boldsymbol{\gamma}_b^{\alpha+1} = 0.5[(1-\boldsymbol{\boldsymbol{\beta}})\boldsymbol{\gamma}_a^\alpha+(1+\boldsymbol{\beta})\boldsymbol{\gamma}_b^\alpha], \label{31}
\end{aligned}
\end{equation}
where \textbf{$\boldsymbol{\gamma}_a^{\alpha+1}$} and \textbf{$\boldsymbol{\gamma}_b^{\alpha+1}$} are the offspring, \textbf{$\boldsymbol{\gamma}_a^{\alpha}$} and \textbf{$\boldsymbol{\gamma}_a^{\alpha}$} are the parents. \textbf{$\boldsymbol{\beta}$} $\in \mathbb{R}^{1\times \lceil \widetilde{M_2}\rceil}$ is a randomly generated vector in each crossover operation, $\beta_m$ is the $m$-th element, defined as

\begin{equation}
\begin{aligned}
   \beta_m=
\begin{cases}
\left( 2\nu_m \right)^{\frac{1}{\eta_c+1}}, & \nu_m\leq 0.5, \\
\left( \frac{1}{2\left( 1 - \nu_m \right)} \right)^{\frac{1}{\eta_c+1}}, & else,
\end{cases}
\end{aligned}
\end{equation}
where $\eta_c$ is the crossover distribution index. $\nu_m \sim \text{U}[0, 1]$ is a uniformly distributed random variable, $m\in \left\{1,2,\dots,V \right\}$. The crossover continues until $pc$ offspring are generated.

During the mutation, $pm$ individuals are randomly selected from $Q_t$, and each selected individual undergoes mutation as  
\begin{equation}
\boldsymbol{\gamma}_a^{\alpha+1} = \boldsymbol{\gamma}_a^\alpha+\boldsymbol{\mu} \times \Gamma\times(\varphi_u-\varphi_l), \label{33}
\end{equation}
where \textbf{$\boldsymbol{\mu}$} $\in \mathbb{R}^{1\times \lceil \widetilde{M_2}\rceil}$ is a randomly generated vector in each mutation operation, $\mu_n$ is the $n$-th element, defined as
\begin{equation}
\mu_n=\left\{
             \begin{array}{lr}
             [2\sigma_n+(1-2\sigma_n)(1-\mu_a)^{\eta_m+1}]^\frac{1}{\eta_m+1}-1,  \sigma_n\leq 0.5, \\
             1-[2(1-\sigma_n)+(2\sigma_n-1)(1-\mu_b)^{\eta_m+1}]^\frac{1}{\eta_m+1},  else,
             \end{array}
\right.
\end{equation}
where $\sigma_n \sim \text{U}[0, 1]$, $n\in \left\{1,2,\dots,V \right\}$, $\eta_m$ is the mutation distribution index. $\mu_a$ and $\mu_b$ are defined as $\mu_a = (\gamma_a^\alpha(i) - \varphi_l)/(\varphi_u - \varphi_l)$ and $\mu_b =  (\varphi_u - \gamma_a^\alpha(i))/(\varphi_u - \varphi_l)$, respectively. $\gamma_a^\alpha(i)$ and $\gamma_b^\alpha(i)$ are the $i$-th element in \textbf{$\boldsymbol{\gamma}_a^{\alpha}$} and \textbf{$\boldsymbol{\gamma}_b^{\alpha}$}. The $\Gamma$ in (\ref{33}) is the mutation probability and is defined as
\begin{equation}
\Gamma=\left\{
             \begin{array}{lr}
             1, & \chi\leq \frac{\eta_m}{\lceil \widetilde{M_2} \rceil}, \\
             0, & else,
             \end{array}
\right.
\end{equation}
where $\chi \sim \text{U}[0, 1]$ is a uniformly distributed random variable selected at each mutation. 
We set the mutation to occur with probability 
$p\left(\chi\leq {\eta_m}/{\lceil \widetilde{M_2} \rceil}\right)$, meaning that the mutation probability increases as the sailing time decreases. This design promotes diversity among individuals, preventing premature convergence to a local optimum.

Considering the high dimensionality of individuals, traditional crossover operations often struggle to conduct effective searches and frequently produce variables that violate steering angle constraints. Repeatedly assigning penalty values to such infeasible individuals is inefficient. To address this issue, we conduct a parent smoothing operation before mutation. For each selected parent individual, the differences between adjacent variables are calculated. If the difference violates the steering angle constraint, the corresponding variable is replaced with the average of its two neighboring variables. This smoothing mechanism helps reduce the generation of infeasible solutions and enhances convergence efficiency.

The offspring generated from crossover and mutation are then combined with the parent population to form the new merged population $\widetilde{Q}^\alpha$.

\textbf{Step 4: Evolution to the next generation.} We perform non-dominated sorting and crowding distance calculation on $\widetilde{Q}^\alpha$. The top $N_p$ individuals are selected from $\widetilde{Q}^\alpha$ to form the next-generation population $Q^{\alpha+1}$, and the unselected individuals are discarded. If the current generation $\alpha$ is smaller than the maximum number of generations $G_{max}$, go to \textit{Step 2}; otherwise, go to \textit{Step 5}.

\textbf{Step 5: PSO particles initialization.} We generate the PSO particles from the NSGA-II population, initializing each particle’s position based on its corresponding individual. The velocity of each particle is randomly initialized within the range $\varrho(\varphi_u - \varphi_l)$, where $\varrho$ is a parameter used to control the particle's velocity.  In addition, the highest dominance level individuals in the NSGA-II population are selected as the expected solution set for the problem \textbf{(P3)}, denoted by $\Re^1$. At the end of each generation of the PSO algorithm, this expected solution set is updated and denoted as $\Re^\rho$ in the $\rho$-th generation.


\textbf{Step 6: Evolution to the next-generation particles $Q_{P}^{\rho+1}$.} For each particle in $(\rho+1)$-th generation, its velocity $v^{\rho+1}$ is updated using (\ref{26}). The weight $\omega$ is dynamically adjusted based on the generation number $\rho$, as follows 
\begin{equation}
\omega = \omega_{1} - \rho \times \left(\frac{\omega_{1} - \omega_{2}}{G_{max}^{'}}\right),
\end{equation}
where \(\omega_{1}\) and \(\omega_{2}\) are the upper and lower bounds of the weight $\omega$, respectively. \( G_{max}^{'} \) denotes the maximum number of generations in the PSO algorithm. For the first generation, $g_{best}$ is set as the position of the best-performing particle among all particles, and $ p_{best}$ is initialized as the particle’s own position. In the $\rho+1$ generation, $ g_{best}$ is set as the position of the first particle in $\Re^{\rho+1}$, and $p_{best} = max( p_{best}, p^{\rho})$. By substituting $v^{\rho+1}$ into (\ref{27}), the updated position of the particle, $p^{\rho+1}$, is obtained. $Q_{P}^{\rho+1}$ is then formed accordingly.               

\textbf{Step 7: The expected solution set $\Re^{\rho+1}$ generation.} All particle positions from $Q_P^{\rho+1}$  are used to generate new individuals, which are then added to $\Re^{\rho}$ to form $\Re^{\rho+1}$. Non-dominated sorting is performed on the updated $\Re^{\rho+1}$, and only the highest dominance level individuals are retained. When $\rho$ is less than the maximum number of generation $G_{max}^{'}$, go to \textit{Step 6}, otherwise, the DPPI-NSGA-II ends, and $\Re^{G^{'}_{max}+1}$ is output as the final result.



\begin{figure}[h]
\centering
\includegraphics[width=0.47\textwidth]{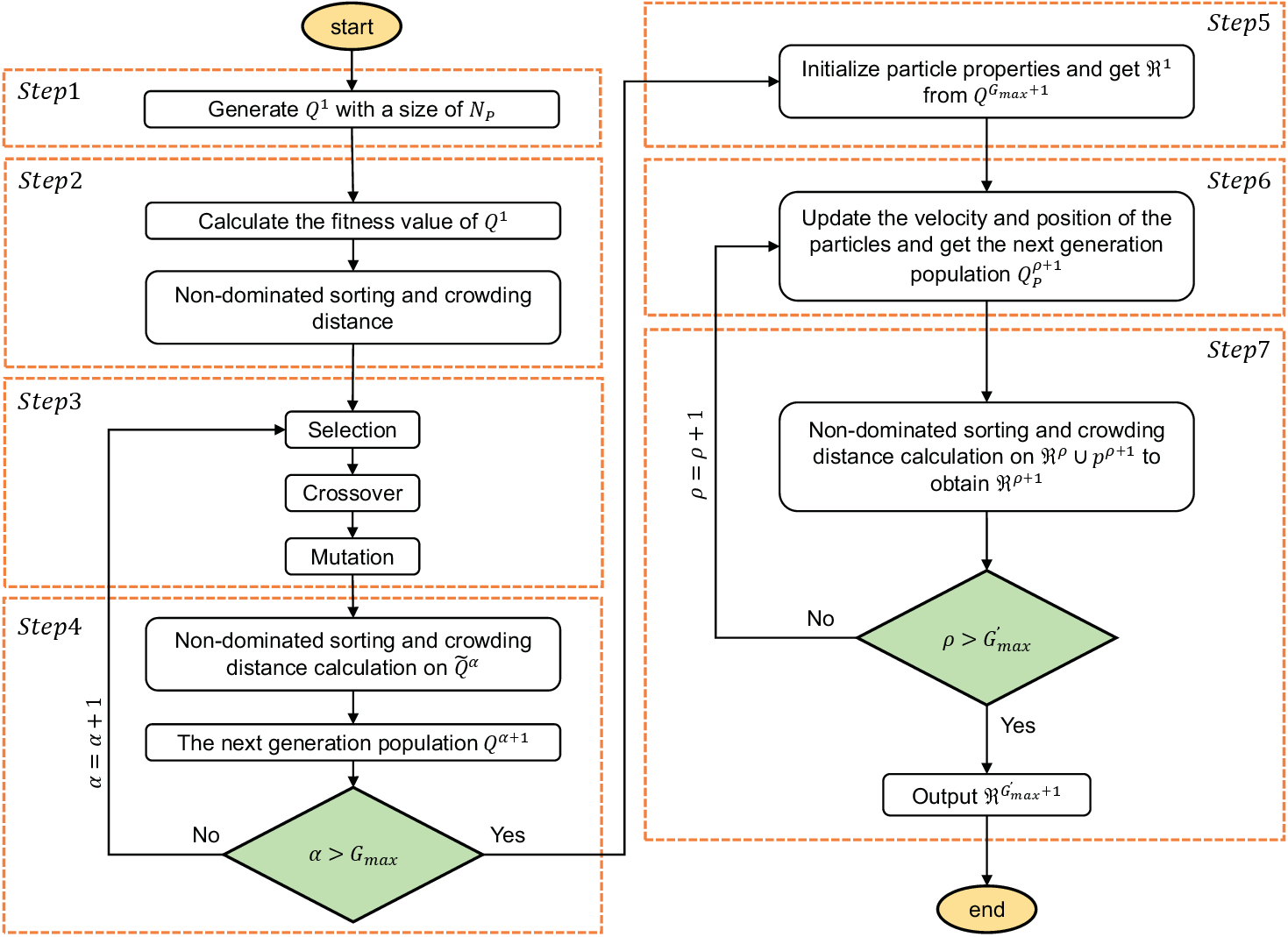}
\captionsetup{font={small,stretch=1.25},justification=raggedright}
\caption{DPPI-NSGA-II algorithm flowchart}
\label{flowchart}
\end{figure}

Overall, the procedure is summarized in Algorithm \ref{alg:AOA}, while Fig.~\ref{flowchart} offers a complementary flowchart illustration.

\begin{algorithm}[!h]
    \caption{DPPI-NSGA-II}
    \label{alg:AOA}
    \begin{algorithmic}[1]
        \State \textbf{Initialization:} $\gamma_i, \forall i \in {1,2,...,N_P}$, $Q^1$
        \For{$\alpha=1$ to $G_{max}$}
            \State Non-dominated sorting (\ref{28})(\ref{29})(\ref{30}) 
            \State $Q_t$ $\leftarrow$ top $pd$ individuals from $Q^\alpha$
            \State Set $Q_{cro} = \emptyset$
            \For{$\forall \gamma_a^{\alpha}, \gamma_b^{\alpha} \in Q_t$}
                \State Crossover (\ref{31}); $Q_{cro} \leftarrow Q_{cro} \cup \gamma_a^{\alpha+1} \cup \gamma_b^{\alpha+1}$
             \EndFor
            \State $Q^{'}_t$ $\leftarrow$ top $pm$ individuals from $Q^\alpha$ 
            \State Set $Q_{mut} = \emptyset$
            \For{$\forall \gamma_a^{\alpha} \in Q^{'}_t$}
                \State Mutation (\ref{33}); $Q_{mut} \leftarrow Q_{mut} \cup \gamma_a^{\alpha+1}$
             \EndFor
            \State $\widetilde{Q}^\alpha \leftarrow Q^\alpha \cup Q_{cro} \cup Q_{mut}$
            \State $Q^{\alpha+1}$ $\leftarrow$ top $N_p$ individuals from $\widetilde{Q}^\alpha$
        \EndFor
        \For{each particle in $Q_P^{1}$ }
            \State $p^1 \leftarrow \gamma_i$; $v^1$ randomly initialized
        \EndFor
        \State $\Re^{1} \leftarrow$ highest dominance level individuals from $Q^{G_{max}+1}$
        \For{$\rho = 1$ to $G^{'}_{max}$}
            \For{each particle in $Q_P^{\rho+1}$}
                \State $v^{\rho+1} \leftarrow (\ref{26})$; $p^{\rho+1} \leftarrow (\ref{27})$
                \State update $p_{best}$ and $g_{best}$
                \State $\Re^{\rho+1} \leftarrow \Re^{\rho} \cup p^{\rho+1}$
            \EndFor
            \State update $\Re^{\rho+1}$ by non-dominated sorting
            \State $\Re^{\rho+1}$ retains the highest dominance level individuals
        \EndFor
        \State \textbf{Output:} $\Re^{G^{'}_{max}+1}$
    \end{algorithmic}
\end{algorithm}            

\subsection{Complexity Analysis}
The computational complexity of DPPI-NSGA-II mainly arises from two components: non-dominated sorting and fitness function calculation. The complexity of non-dominated sorting is $O(N_{obj}(N_p+pc+pm)^2)$, where $N_{obj}$ is the number of optimization objectives, $N_p$ is the population size, and $pc$ and $pm$ are the number of offspring generated by crossover and mutation, respectively. The fitness function calculation involves two steps: CGM's grid division and data transmission calculation, as defined in (\ref{25c}). Assuming a constant solution space dimension in the worst case, the complexity of the fitness function calculation is $O(N_pT/\delta t)$. The expected solution set update in each generation involves non-dominated sorting and crowding distance calculation, with the sorting complexity becoming $O(N_{obj}(N_p)^2)$. Combining all components, the overall computational complexity of the DPPI-NSGA-II algorithm is expressed as $O((G_{max}+G_{max}^{'})(N_{obj}(N_p)^2 + N_p T/\delta t) $, where \(G_{max}\) and \(G_{max}^{'}\) denote the maximum number of generations in NSGA-II and PSO, respectively.

\section{Simulation Results}
\label{section:D}

\begin{figure*}[hbtp]
	\centering
	\begin{subfigure}{0.32\linewidth}
		\centering
		\includegraphics[width=1.1\linewidth]{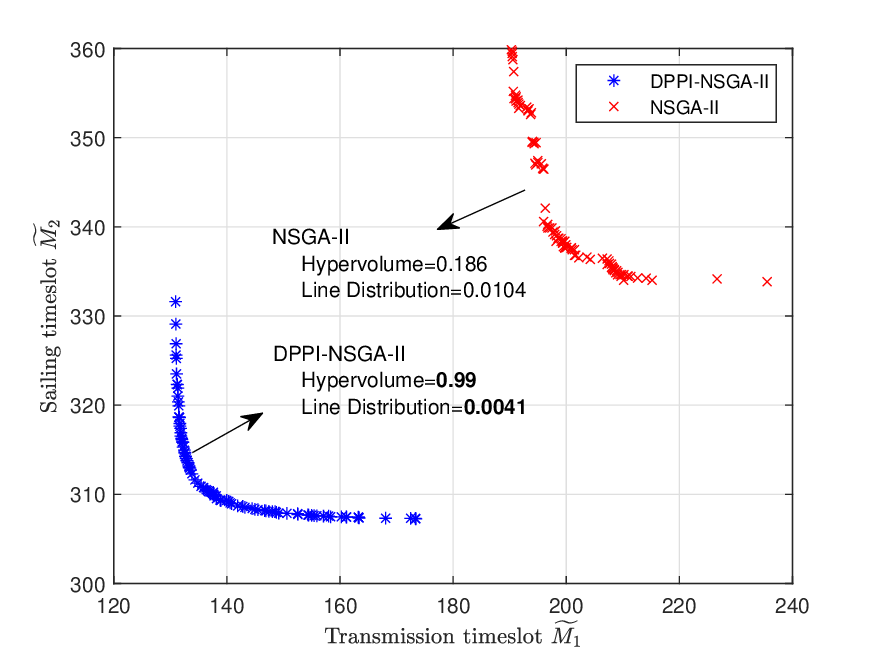}
		\caption{\footnotesize Case 1:$X_A[-50,50]$~km, $X_B[70,70]$~km}
		\label{duct1}
	\end{subfigure}
	\centering
	\begin{subfigure}{0.32\linewidth}
		\centering
		\includegraphics[width=1.1\linewidth]{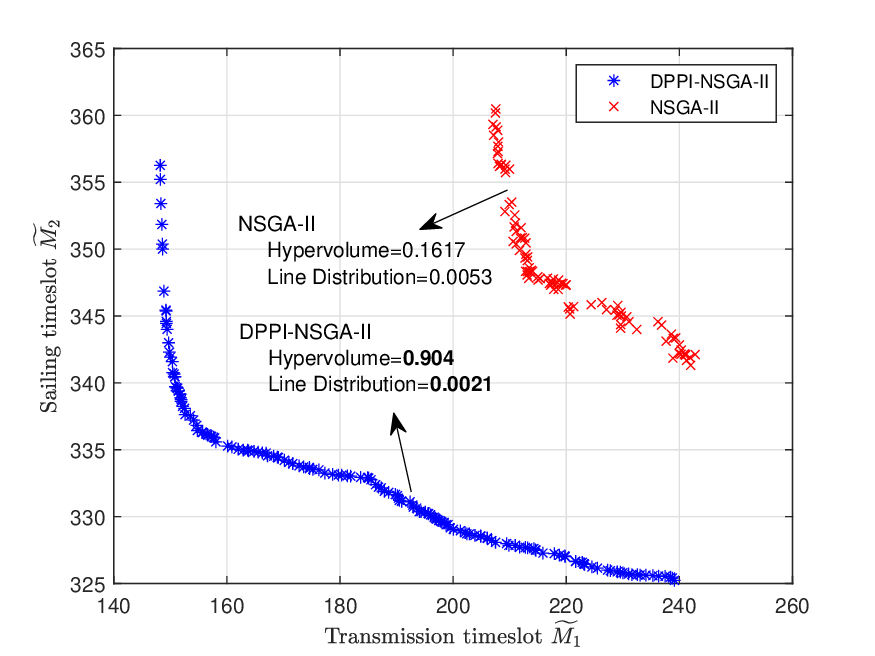}
		\caption{\footnotesize Case 2:$X_A[-70,70]$~km, $X_B[50,50]$~km}
		\label{duct2}
	\end{subfigure}
	\centering
	\begin{subfigure}{0.32\linewidth}
		\centering
		\includegraphics[width=1.1\linewidth]{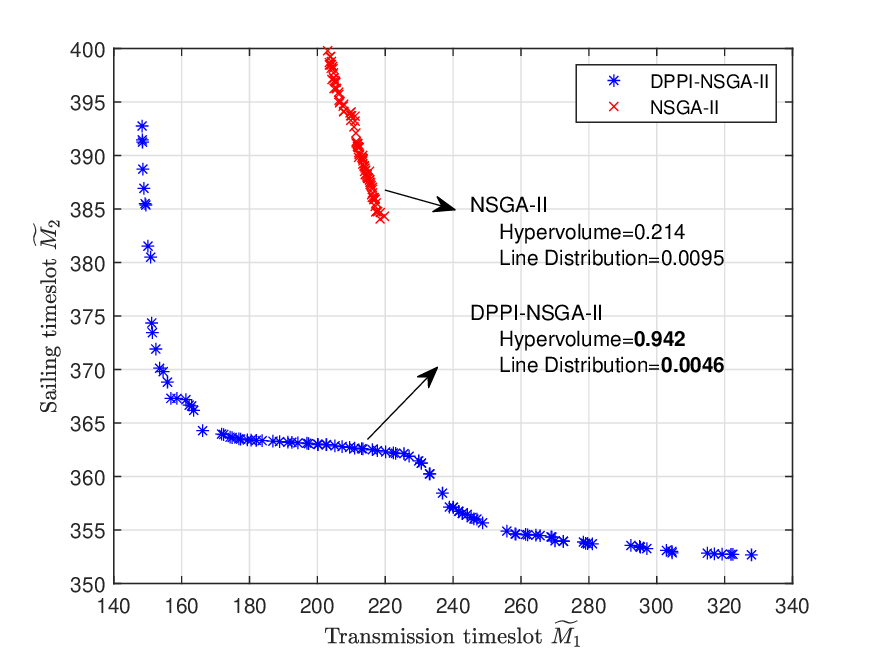}
		\caption{\footnotesize Case 3:$X_A[-70,70]$~km, $X_B[70,70]$~km}
		\label{duct3}
	\end{subfigure}
    \caption{Comparison of Pareto front obtained from NSGA-II and DPI-NSGA-II algorithms under three different cases.}
	\label{CASE}
\end{figure*}

In this section, we conduct simulations to demonstrate the effectiveness of the proposed scheme. In the simulation, we use PETOOL based on the PE method \cite{ozgun2011petool} to generate the path loss of the CGM, assuming a uniformly distributed evaporation duct with a constant height $edh = 35$ m. The horizontal and vertical resolutions of the CGM are set to $\Delta d = 50$ m and $\Delta h = 1$ m, respectively. The carrier frequency $f$ is set to 10 GHz and the bandwidth $B$ is 50 MHz \cite{saafi2023cost}. The transmit antenna height and receive antenna height are set to $z_{tx} = 10$ m and $z_{rx} = 15$ m. The maximum LoS transmission distance $d_{\mathrm{LoS}}$ is calculated to be 29 km \footnote{$d_{\mathrm{LoS}}$ is calculated from $ 4.12 \times (\sqrt{z_{tx}}+\sqrt{z_{rx}})$, where the factor 4.12 is a correction factor to accommodate standard atmospheric refraction conditions \cite{deminco2007propagation}.}. Transmitter and receiver antenna gains are set to 15 dBi and 20 dBi, respectively; the transmit power is 15 dBm, and the noise power spectral density is -169 dBm/Hz \cite{zhang2024field}. The speed of the SU $v$ is fixed at 20 m/s and the timeslot duration is set to $\Delta t = 20$ s. The corresponding maximum steering angle per timeslot is set to $\Delta_{\varphi_{\text{max}}} = \pi/4$ \cite{neatby2024turning}. All the start and end locations are assumed to be located beyond the LoS transmission range. The following three scenarios are considered.


\begin{itemize}
    \item \textbf{Case 1}: Far end location scenario, $X_A = [-50,50]$~km, $X_B = [70,70]$~km;
    \item \textbf{Case 2}: Far start location scenario, $X_A = [-70,70]$~km, $X_B = [50,50]$~km;
    \item \textbf{Case 3}: Far start and end locations scenario, $X_A = [-70,70]$~km, $X_B = [70,70]$~km.
\end{itemize}

The detailed simulation parameters are provided in Table \ref{tab:example2}. We first compare the performance of the NSGA-II and DPPI-NSGA-II algorithms from multiple perspectives. Next, we analyze the transmission timeslots $\widetilde{M_1}$ and sailing timeslots $\widetilde{M_2}$, both with and without the CGM, under evaporation duct conditions. The SU’s sailing trajectory and the corresponding variation in data transmission are also examined. 

\begin{center}
\begin{footnotesize}
\renewcommand{\arraystretch}{1.1} 
\setlength{\tabcolsep}{4pt} 
\captionof{table}{Simulation parameters}
\label{tab:example2}
\begin{tabularx}{0.45\textwidth}{X|X|X|X} 
 \hline
 \textbf{Parameter} & \textbf{Value} & \textbf{Parameter} & \textbf{Value} \\
 \hline
 \hline
 $edh$         & 35 m            & $D$          &  40 GB          \\ 
 $\Delta t$         & 20 s            & $v$               & 20 m/s          \\
  $\Delta d$         & 50 m            & $\Delta h$               & 1 m          \\
 $\varphi_{\max}$   & $\pi/4$        & $z_{\text{tx}}$   & 10 m            \\
 $z_{\text{rx}}$    & 15 m            & $P_t$             & 15 dBm          \\
 $G_t$              & 15 dBi          & $G_r$             & 20 dBi          \\
 $B$                & 50 MHz          & $f$               & 10 GHz         \\
 $n_0$              & -169 dBm/Hz     &                &          \\
 \hline
 \end{tabularx}
\end{footnotesize}
\end{center}

\subsection{Algorithm Performance}
The performance of the proposed DPPI-NSGA-II algorithm is evaluated under three cases. The traditional NSGA-II algorithm is selected as the benchmark for comparison. Both algorithms are initialized with the same population size and executed under identical runtime conditions to solve problem P3. Fig.~\ref{CASE} shows the resulting Pareto fronts, where the horizontal axis represents the transmission timeslot $\widetilde{M_1}$ and the vertical axis represents the sailing timeslot $\widetilde{M_2}$. From Fig.~\ref{CASE}, it can be observed that in all three cases, the proposed DPPI-NSGA-II algorithm outperforms the traditional NSGA-II algorithm. For example, in Case 1, the DPPI-NSGA-II can obtain transmission timeslots of the lowest 131 and sailing timeslots of 307, while the NSGA-II is 190 and 307, respectively. To further compare the performance of the algorithms, we introduce the hypervolume  \cite{liu2015multi} and line distribution  \cite{ibrahim20183d} to evaluate the quality of solution sets from the Pareto fronts.


The hypervolume quantifies the multidimensional volume occupied by the solution set relative to a predefined reference point in the objective space. A larger hypervolume value indicates better convergence and diversity of the Pareto front. Its essence is to calculate the volume of the area dominated by the solution set in the objective space. Let $HV$ denote the hypervolume, which can be obtained from 
\begin{equation}
HV(S,\omega) = \Lambda(\bigcup \limits_{S} [f^1,\omega^1]\times [f^2,\omega^2]\times \cdots \times [f^m,\omega^m]),
\end{equation}
where $S=[f^1,f^2,\cdots,f^m]$ is the set of Pareto solutions, $m$ is the number of solutions and each $f^i$ is a Pareto solution. $\Lambda$ is the Lebesgue measure, which is the standard measure on Euclidean space. $\omega = [\omega^1,\omega^2,\cdots,\omega^m]$ is the reference point.  $[f^1,\omega^1]\times [f^2,\omega^2]\times\cdots\times [f^m,\omega^m]$ represents the hypercuboid formed by the solution set $S$.

We normalized the solution sets obtained by the proposed DPPI-NSGA-II and the traditional NSGA-II algorithms by using the hypervolume value enclosed by the reference and the minimum value in the solution set as the unit hypercube. We marked the calculated hypervolume values in Fig.~\ref{CASE}. The results show that the hypervolume values of the Pareto fronts obtained by DPPI-NSGA-II are greater than those achieved by NSGA-II across all three cases. This indicates that DPPI-NSGA-II exhibits better convergence performance, with its Pareto fronts being closer to the true Pareto front.
    
The line distribution evaluates the scalability and uniformity of the Pareto solution set. For a given objective’s target interval, partition it into $m$ equal subintervals, compute the mean distance from each segment midpoint to the nearest solution value on that objective. Smaller values indicate more uniform coverage. Let $\Delta_{Line}^i$ denote the line distribution of objective $i$, which can be obtained from
\begin{equation}
\Delta_{Line}^i(S) = \frac{\sum_{j=1}^{|\zeta|} \mathop{\text{min}}\limits_{s\in S}|\zeta_j-F_i(s)| }{|\zeta|},
\end{equation}
Where $F_i(s)$ is the normalized value of the $i$-th solution. 
$\zeta_j$ denotes the midpoint of the $i$-th interval between two consecutive solutions, and $|\zeta|$ represents the total number of intervals. The overall line distribution $\Delta_{Line}(S)$ can be obtained from
\begin{equation}
\Delta_{Line}(S) = \frac{\sum_{i=1}^{N_{obj}}\Delta_{Line}^i(S) }{N_{obj}},
\end{equation}
where $N_{obj}$ is the number of objectives.
    
We calculate the $\Delta_{Line}$ for both algorithms, and the results are shown in Fig.~\ref{CASE}. The results show that the $\Delta_{Line}$ is consistently lower for DPPI-NSGA-II across all three cases, indicating better solution uniformity and distribution compared to NSGA-II.

\begin{figure}[h!]
\centering
\includegraphics[width=0.45\textwidth]{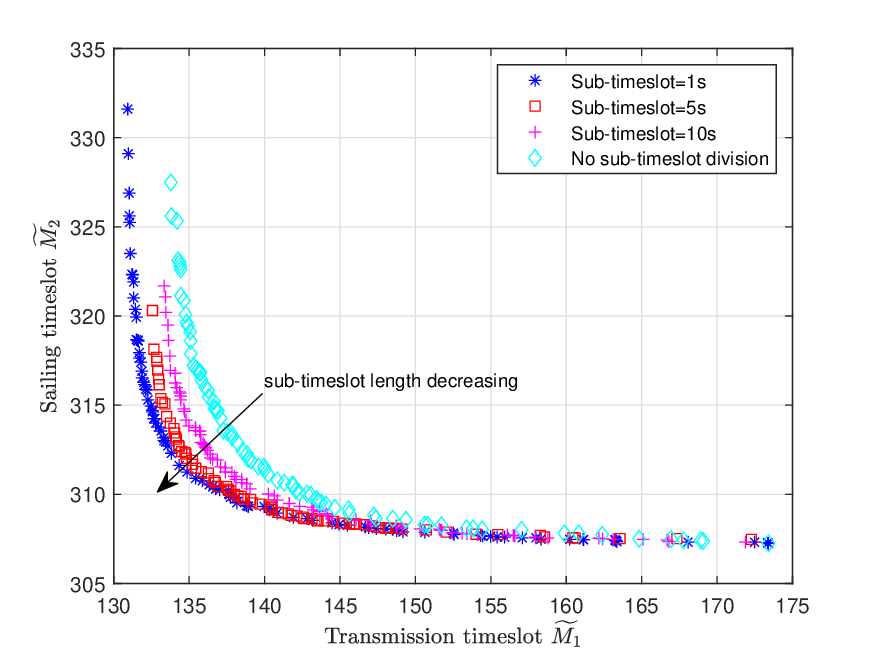}
\captionsetup{font={small,stretch=1.25},justification=raggedright}
\caption{Comparison of Pareto front with sub-timeslots of $\delta t=1$~s, $\delta t=5$~s, $\delta t=10$~s and no sub-timeslots division.}
\label{fig161}
\end{figure}

\begin{figure*}[h!]
	\centering
	\begin{subfigure}{0.32\linewidth}
		\centering
		\includegraphics[width=1.1\linewidth]{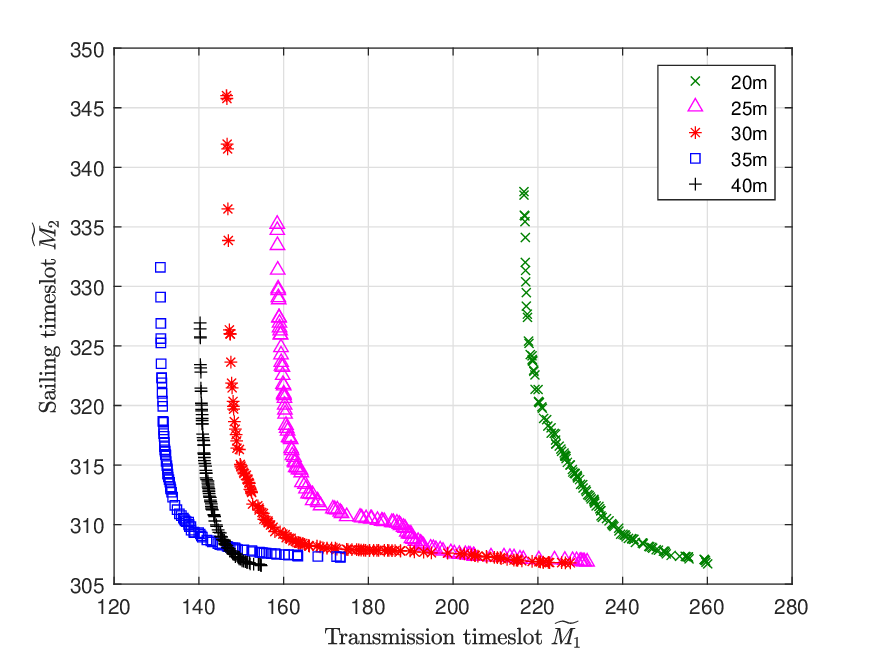}
		\caption{\footnotesize Case 1:$X_A[-50,50]$~km, $X_B[70,70]$~km}
		\label{duct1}
	\end{subfigure}
	\centering
	\begin{subfigure}{0.32\linewidth}
		\centering
		\includegraphics[width=1.1\linewidth]{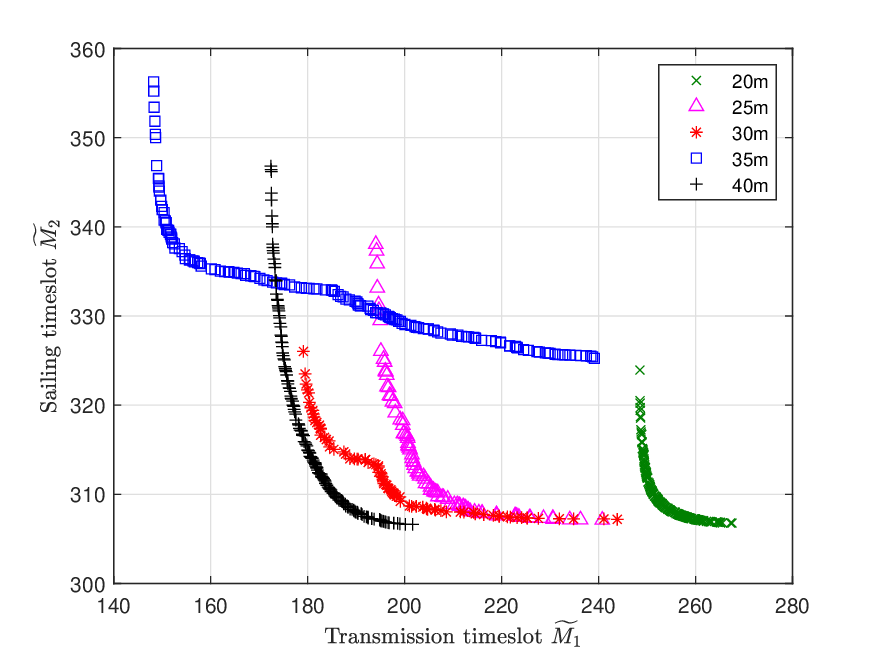}
		\caption{\footnotesize Case 2:$X_A[-70,70]$~km, $X_B[50,50]$~km}
		\label{duct2}
	\end{subfigure}
	\centering
	\begin{subfigure}{0.32\linewidth}
		\centering
		\includegraphics[width=1.1\linewidth]{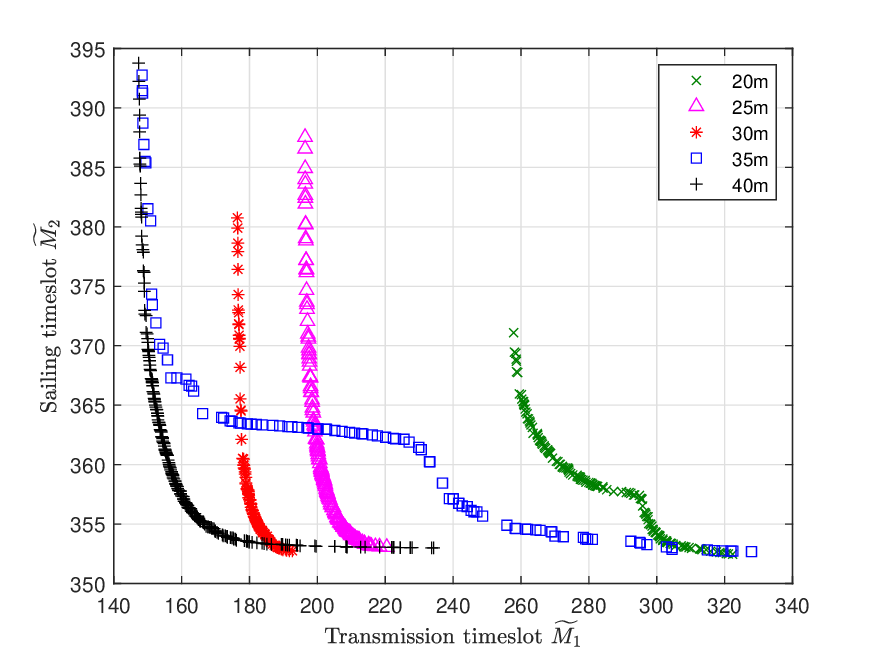}
		\caption{\footnotesize Case 3:$X_A[-70,70]$~km, $X_B[70,70]$~km}
		\label{duct3}
	\end{subfigure}
    \caption{Comparison of Pareto fronts with and without CGM schemes under three cases, highlighting one representative solution in terms of transmission and sailing timeslots.}
	\label{EDUCASE}
\end{figure*}

Taking Case 1 as an example, we conduct simulations using different sub-timeslot lengths. As shown in Fig.~\ref{fig161}, the Pareto front moves down as the sub-timeslot length decreases, indicating that smaller sub-timeslots lead to reduced approximation errors and more accurate results. However, the difference between the results becomes much smaller when $\delta t=5$~s and $\delta t=1$~s. This suggests that once the sub-timeslot length is sufficiently small, further reductions have a negligible impact on results.

We further examined the Pareto front performance for Cases 1, 2, and 3 at different evaporation duct heights, as shown in Fig.~\ref{EDUCASE}. It can be observed that the evaporation duct height significantly impacts both sailing time and data transmission time. Under the same sailing time, the transmission times vary considerably across different duct height conditions, yet do not decrease monotonically with increasing height. This indicates that the relationship between duct height and transmission efficiency is non-linear, possibly due to the complex interplay between duct height, wave trapping efficiency, and reflection loss. Similar trends are also observed under the same transmission time. Furthermore, the Pareto fronts of the same evaporation duct height in different cases vary greatly, which indicates that the starting and ending points also have an impact on trajectory optimization. These findings verify that evaporation duct height has a significant effect on both trajectory and transmission performance, leading to the importance of designing communication trajectories based on environment-specific CGM. 

\begin{figure}[h!]
\centering
\includegraphics[width=0.45\textwidth]{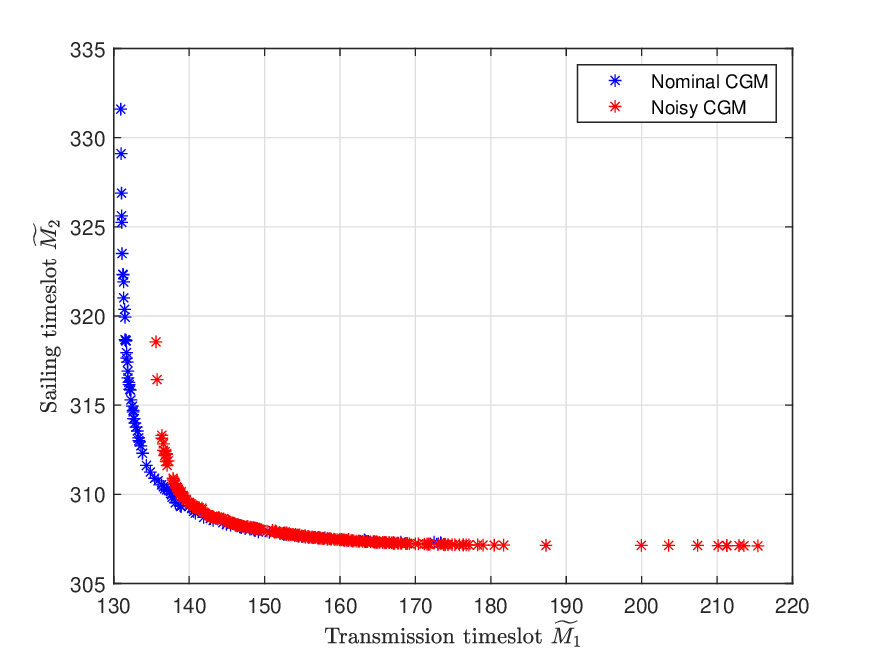}
\captionsetup{font={small,stretch=1.25},justification=raggedright}
\caption{Impact of CGM noise on Pareto front performance.}
\label{errors}
\end{figure}

We also evaluated the potential impact of CGM noise on algorithm performance. To assess robustness, we generated a noisy CGM by adding independent zero-mean Gaussian noise to every grid cell. Using this noisy CGM, we re-ran the Case~1 trajectory optimization under identical settings and compared the resulting Pareto fronts with those from the nominal map. As shown in Fig.~\ref{errors}, noise induces a slight degradation of the Pareto front. The loss arises when spurious high-gain patches steer trajectory segments into genuinely low-gain regions and increase attenuation. The algorithm remained stable and converged normally, and no functional failures were observed, indicating robustness to practical CGM uncertainties.
 
\begin{figure*}[h!]
	\centering
	\begin{subfigure}{0.32\linewidth}
		\centering
		\includegraphics[width=1.1\linewidth]{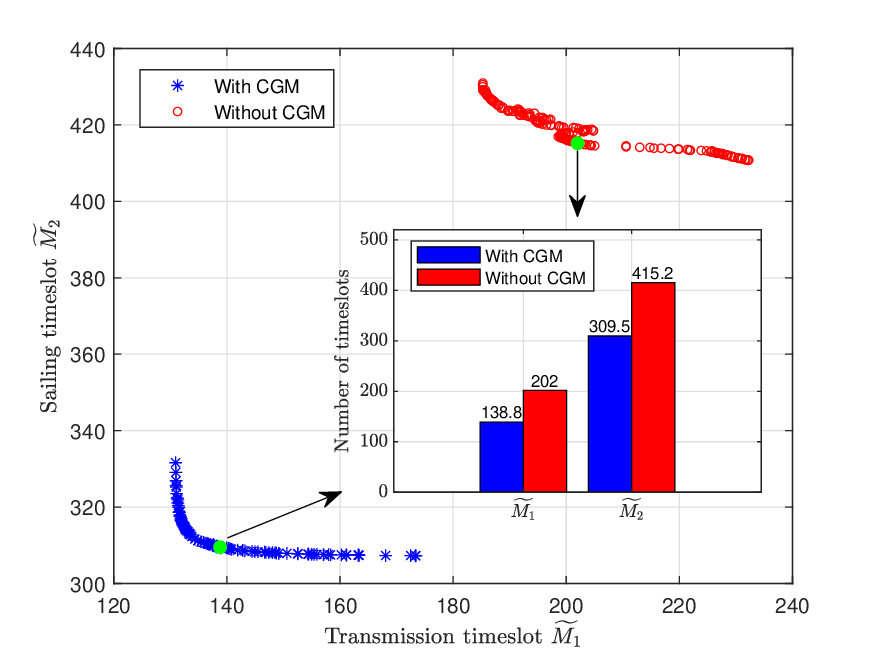}
		\caption{\footnotesize Case 1:$X_A[-50,50]$~km, $X_B[70,70]$~km}
		\label{duct1}
	\end{subfigure}
	\centering
	\begin{subfigure}{0.32\linewidth}
		\centering
		\includegraphics[width=1.1\linewidth]{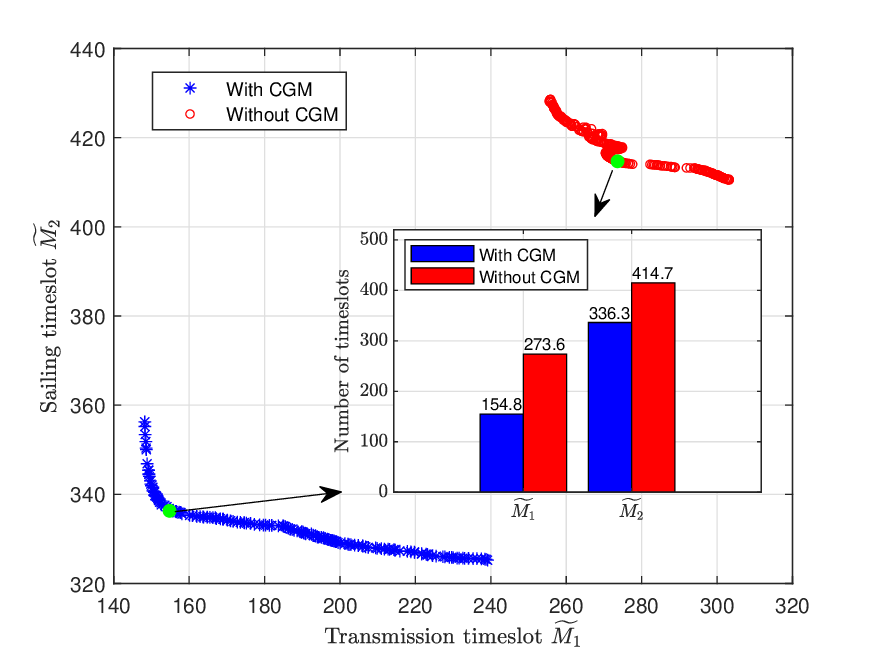}
		\caption{\footnotesize Case 2:$X_A[-70,70]$~km, $X_B[50,50]$~km}
		\label{duct2}
	\end{subfigure}
	\centering
	\begin{subfigure}{0.32\linewidth}
		\centering
		\includegraphics[width=1.1\linewidth]{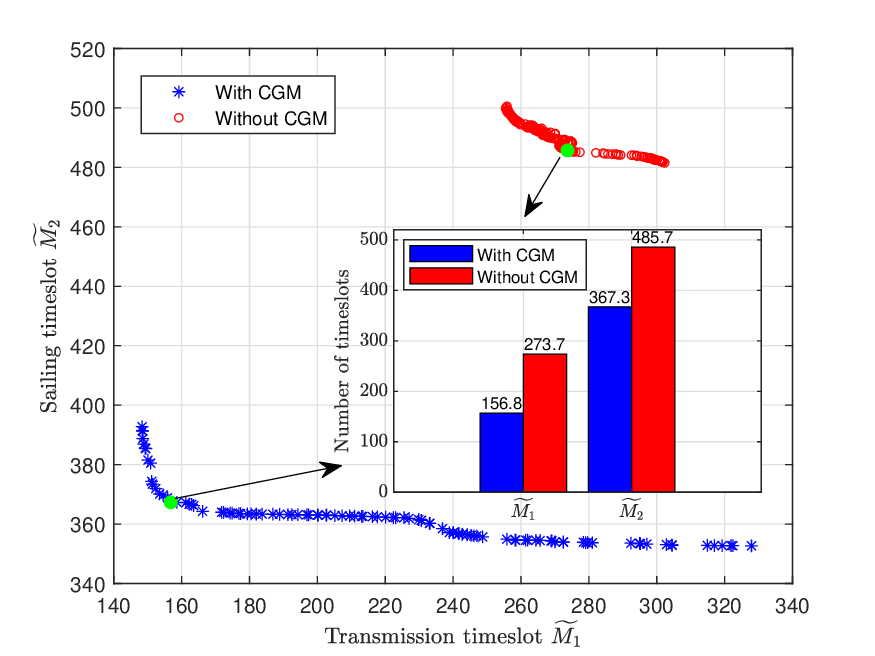}
		\caption{\footnotesize Case 3:$X_A[-70,70]$~km, $X_B[70,70]$~km}
		\label{duct3}
	\end{subfigure}
    \caption{ Comparison of Pareto fronts with and without CGM schemes under three cases, highlighting one representative solution in terms of transmission and sailing timeslots.}
	\label{CASE1}
\end{figure*}

\begin{figure*}[h!]
	\centering
	\begin{subfigure}{0.32\linewidth}
		\centering
		\includegraphics[width=1.1\linewidth]{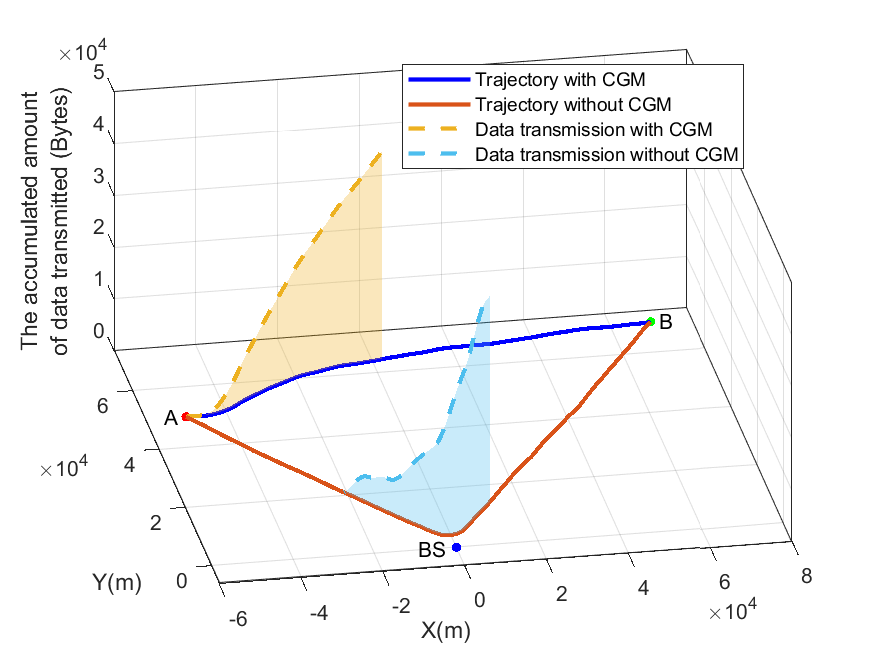}
		\caption{\footnotesize Case 1:$X_A[-50,50]$~km, $X_B[70,70]$~km\\ \footnotesize With CGM:$\widetilde{M_1} = 138.8$, $\widetilde{M_2} = 309.5$\\ \footnotesize Without CGM:$\widetilde{M_1} = 202$, $\widetilde{M_2} = 415.2$}
		\label{duct1}
	\end{subfigure}
	\centering
	\begin{subfigure}{0.32\linewidth}
		\centering
        \includegraphics[width=1.1\linewidth]{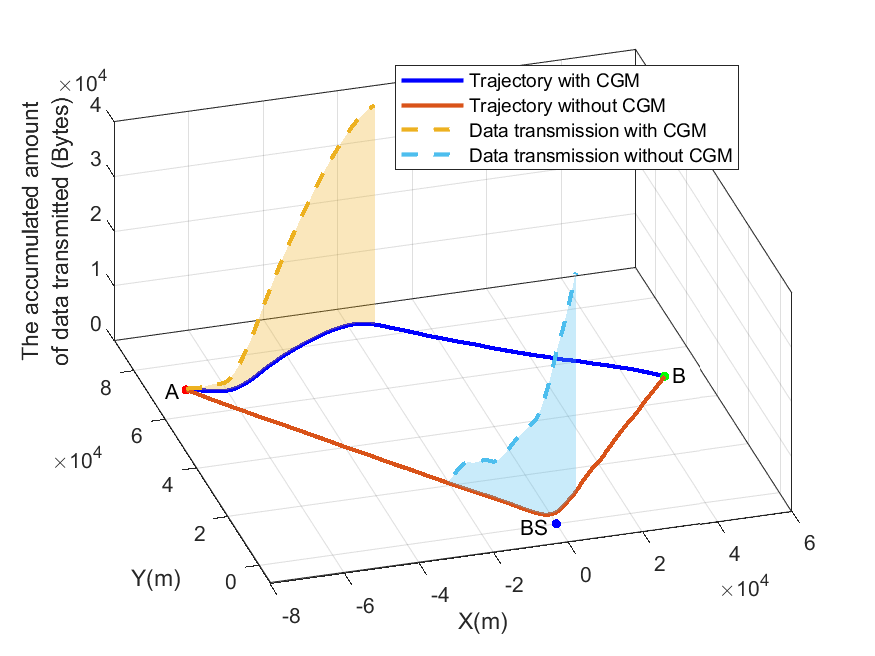}
		\caption{\footnotesize Case 2:$X_A[-70,70]$~km, $X_B[50,50]$~km\\ \footnotesize With CGM:$\widetilde{M_1} = 154.8$, $\widetilde{M_2} = 336.3$\\ \footnotesize Without CGM:$\widetilde{M_1} = 273.6$, $\widetilde{M_2} = 414.7$}
		\label{duct2}
	\end{subfigure}
	\centering
	\begin{subfigure}{0.32\linewidth}
		\centering
		\includegraphics[width=1.1\linewidth]{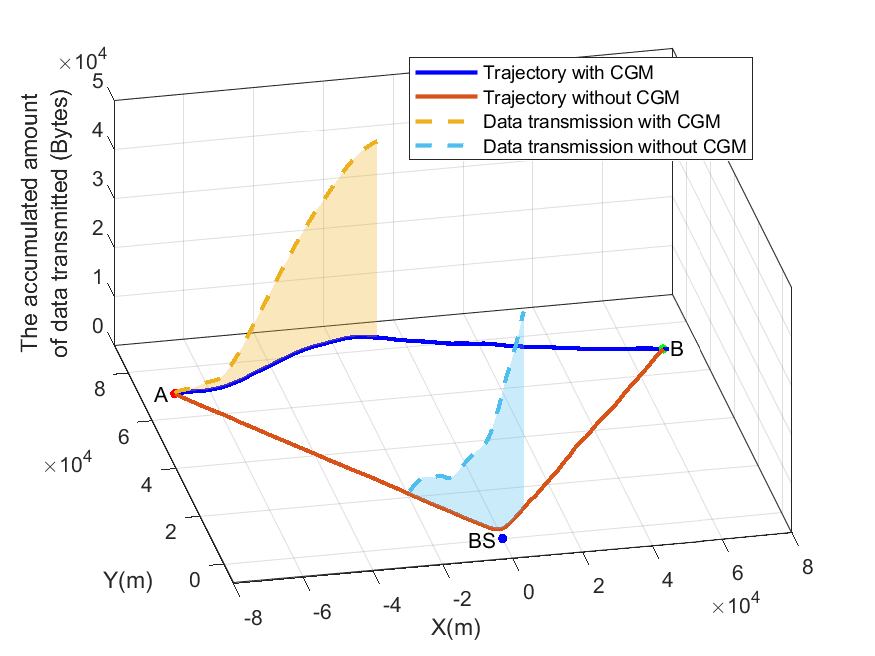}
		\caption{\footnotesize Case 3:$X_A[-70,70]$~km, $X_B[70,70]$~km\\ \footnotesize With CGM:$\widetilde{M_1} = 156.8$, $\widetilde{M_2} = 367.3$\\ \footnotesize Without CGM:$\widetilde{M_1} = 273.7$, $\widetilde{M_2} = 485.7$}
		\label{duct3}
	\end{subfigure}
    \caption{SU trajectory and data transmission with and without CGM schemes under three cases.}
	\label{fig:fi93}
\end{figure*}

\subsection{Trajectory and Transmission Performance Using CGM}

In this subsection, we compare the trajectory and transmission performance with and without CGM in the presence of the evaporation duct. In the scheme without CGM, the trajectory and transmission design are based on the free-space path loss model assumption, and the according \textbf{(P3)} is also formulated and solved using the proposed DPI-NSGA-II algorithm. All three cases are considered in the simulations. As shown in Fig.~\ref{CASE1}, the Pareto fronts obtained with CGM are significantly lower than those without CGM. To illustrate this in more detail, we select one representative solution from each Pareto front and present it as a subfigure in each case. The results show that the sailing timeslots under the scheme with CGM are significantly fewer than those without CGM, while the corresponding transmission timeslots are also reduced.

It can also be observed that the number of solutions under the with CGM scheme is greater than that of the scheme without CGM. This is likely because, without CGM, transmission can only occur when the SU is within the LoS range of the BS, forcing the SU to first approach the BS before initiating communication. In contrast, with CGM, the SU uses the evaporation duct information provided by the CGM to enable BLoS communication. This demonstrates the advantage of utilizing the evaporation duct in communications.




Furthermore, we select one representative solution from each case and plot its corresponding trajectory and data transmission process. For comparison, we also consider the scheme without CGM, in which the trajectory is also generated using the proposed DPPI-NSGA-II algorithm. However, under the scheme without CGM, data transmission only begins when the SU sails into the LoS range of the BS. The SU's transmission rate is assumed to be the maximum achievable rate. As shown in Fig.~\ref{fig:fi93}, the X-Y plane represents the sailing trajectory of SU, while the Z-axis indicates the accumulated amount of data transmitted along the trajectory. It can be observed that, under the without CGM scheme, the SU initially sails directly toward the BS. Upon reaching the vicinity of the BS, it then turns to the end point B. In contrast, under the with CGM scheme, the sailing distance is significantly shorter, and the trajectory differs markedly. The SU may even move away from the BS during transmission. This behavior is attributed to the characteristics of evaporation duct propagation, as illustrated in Fig.~\ref{fig:fig3}. The path loss does not increase monotonically with distance but instead exhibits oscillatory behavior. As a result, in some parts of the trajectory, even when the SU is moving away from the BS, the path loss can decrease, thereby enhancing data transmission performance. 

Additionally, it can be observed that under the with CGM scheme, data transmission remains feasible even at distances up to $\sqrt{70^2+70^2}\approx100$ km. Following a brief trajectory adjustment, the transmission rate quickly reaches a relatively high level. In contrast, under the without CGM scheme, the SU is only able to transmit data in the vicinity of the BS, i.e., within the LoS range. Consequently, a substantial amount of time is spent sailing into the LoS range of the BS. These results demonstrate that exploiting evaporation duct propagation can significantly improve communication efficiency.

\subsection{Multi-waypoint trajectory optimization}

\begin{figure}[h!]
\centering
\includegraphics[width=0.45\textwidth]{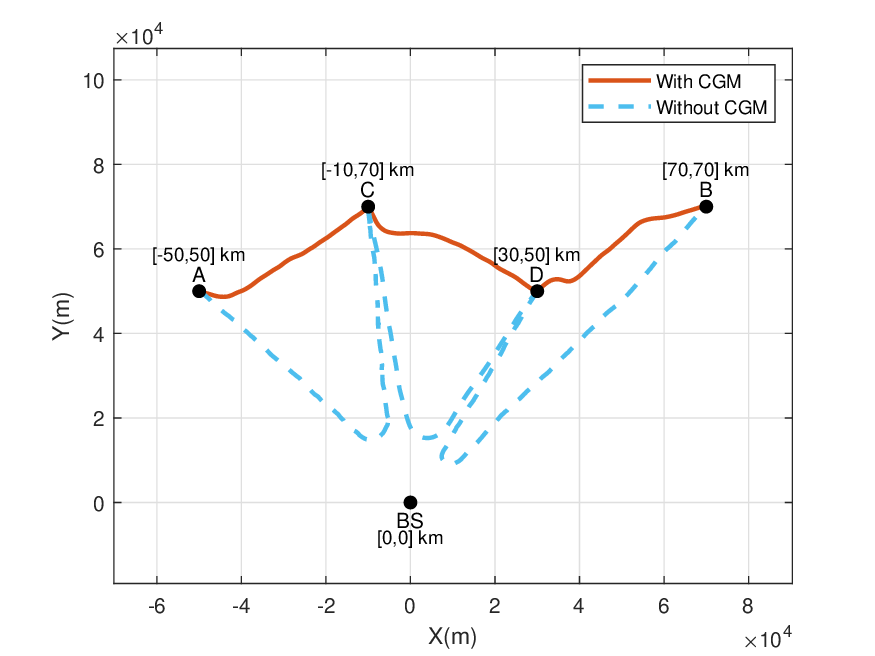}
\captionsetup{font={small,stretch=1.25},justification=raggedright}
\caption{SU trajectory under multi-waypoint scenario where $X_A = [-50,50]$~km, $X_B = [70,70]$~km, $X_C = [-10,70]$~km and $X_D = [30,50]$~km.}
\label{fig8}
\end{figure}

In this subsection, we extend the proposed DPPI-NSGA-II algorithm to a multi-waypoint scenario, where the SU sails from $A$ to $B$ via waypoints $C$ and $D$. The SU collects data at $A$, transmits it to the BS while sailing to $C$, then repeats the process at $C$ and $D$. The overall trajectory is divided into segments $A\rightarrow C$, $C\rightarrow D$, and $D\rightarrow B$, each optimized using DPPI-NSGA-II algorithm. A solution from the Pareto front is shown in Fig.~\ref{fig8}. As shown, without exploiting the information of CGM, the SU must frequently approach the BS to transmit collected data. In contrast, our proposed scheme significantly shortens the sailing distance, and hence greatly improves the efficiency of data collection and transmission.

\section{Conclusion}
\label{section:E}
In this paper, we have investigated the optimization of evaporation duct-based maritime wireless communication by jointly considering channel mapping and trajectory design. A maritime communication model was developed that considered the effects of evaporation ducts and ship navigation constraints, and a simplified motion model was introduced to facilitate trajectory planning. An alignment scheme was proposed to address the mismatch between the resolution of CGM and the ship’s trajectory. We formulated a multi-objective optimization problem and developed a DPPI-NSGA-II algorithm to solve it efficiently. We have shown that our proposed algorithm can improve performance by reducing data transmission and sailing time. With CGM, the SU gains greater flexibility in both communication and sailing, enabling BLoS communication capabilities. As maritime communication continues to gain importance in future network scenarios, we believe that the integration of environmental modeling and intelligent trajectory design offers promising potential for practical deployment.
\printbibliography[title={References}]

\end{document}